\documentclass{PoS}
\usepackage{bm}
\usepackage{amssymb}
\usepackage{fontenc}
\usepackage{times}
\usepackage{mathptmx}
\usepackage{graphicx}
\usepackage{epsfig}
\title{The determination of $\alpha_s(M_Z)$ from perturbative analyses 
of short-distance-sensitive lattice QCD observables revisited }

\ShortTitle{$\alpha_s(M_Z)$ from short-distance-sensitive lattice observables}

\author{\speaker{Kim Maltman}$^a$, D. Leinweber$^b$, P. Moran$^b$ and
A. Sternbeck$^b$%
\\
\llap{$^a$}Mathematics and Statistics, York University, Toronto, Canada\\
\llap{$^b$}CSSM, University of Adelaide, Adelaide, Australia\\
        E-mail: \email{kmaltman@yorku.ca}, 
\email{dleinweb@physics.adelaide.edu.au}, \email{peter.moran@adelaide.edu.au},
\email{andre.sternbeck@adelaide.edu.au}}

\abstract{The determination of $\alpha_s(M_Z)$ via perturbative
analyses of short-distance-sensitive lattice observables is revisited, 
incorporating new lattice data and performing a modified version of
the original analysis. The analysis employs two high-intrinsic-scale 
observables, $\log (W_{11})$ and $\log(W_{12})$, and one lower-intrinsic-scale
observable, $\log(W_{12}/u_0^6)$. We find good consistency among
the values extracted using the different observables and a final result,
$\alpha_s(M_Z)=0.1192\pm 0.0011$, in excellent agreement
with various recent non-lattice determinations, as well as with the
results of a similar, but not identical, re-analysis by the HPQCD 
collaboration. The relation between the two re-analyses is discussed,
focussing on the complementarity of the two approaches.}

\FullConference{The XXVI International Symposium on Lattice Field Theory \\
		 July 14 - 19, 2008\\
		 Williamsburg, Virginia, USA}

\begin{document}
\section{Introduction and Background}
The $n_f=5$ QCD coupling in the $\overline{MS}$ scheme at the 
conventionally defined reference scale $\mu =M_Z$ represents
one of the fundamental parameters of the Standard Model.
The central value of the 2008 PDG assessment~\cite{pdg08qcdreview}, 
$\alpha_s(M_Z)=0.1176(20)$, remains strongly influenced by the 
high-precision lattice result, $\alpha_s(M_Z)=0.1170(12)$, 
obtained in Ref.~\cite{latticealphas} from an analysis of UV-sensitive
lattice observables using the MILC $a\sim 0.18,\, 0.12,$
and $0.09$ fm ensembles. In the last year, a number of
independent determinations have appeared, from a number of different sources, 
yielding typically somewhat higher values. Specifically,
\begin{itemize}
\item updates of the global EW fit, taking into account the new
$5$-loop result for the dimension $0$ OPE contributions~\cite{bck08},
yield $\alpha_s(M_Z)=0.1191(27)$~\cite{bck08,davieretal08};
\item the most recent hadronic $\tau$ decay determinations~\cite{jb08,my08}, 
whose ranges are encompassed by results of Ref.~\cite{my08}, 
yield $\alpha_s(M_Z)=0.1187(16)$ ($\sim 2\sigma$
lower than the earlier result of Ref.~\cite{davieretal08}, for the reasons
discussed in Refs.~\cite{jb08,my08}); 
\item the determination based on $2$ to $10.6$ GeV $R_{had}$ values 
yields $\alpha_s(M_Z)=0.1190\left({}^{+90}_{-110}\right)$~\cite{kst07}; 
\item an update
of the determination from ${\frac{\Gamma [\Upsilon (1s)\rightarrow\gamma X]}
{\Gamma [\Upsilon (1s)\rightarrow X]}}$ yields $\alpha_s(M_Z)=
0.1190\left({}^{+60}_{-50}\right)$~\cite{brambillaetalupsilon07}; 
\item recent determinations, 
$\alpha_s(M_Z)=0.1198(32)$~\cite{alphasherajetcrosssections},
$0.1182(45)$~\cite{alphash1jetshighqsq08}, 
$0.1240(33)$~\cite{alphaseventshapes08} and
$0.1172(22)$~\cite{bs08}, using shape observables in DIS and 
$e^+e^-\rightarrow hadrons$, if averaged naively,
yield a combined result $\alpha_s(M_Z)=0.1193(15)$. 
\end{itemize}
In view of these results, and, in addition, the availability of the new MILC
ensembles with $a\sim 0.15$ and $0.06$ fm, it is timely to revisit
the earlier lattice analysis. We do so by focussing on three observables,
$\log(W_{11})$, $\log(W_{12})$ and $\log(W_{12}/u_0^6)$ which can be
convincingly argued to receive only small non-perturbative contributions
at the scales of the lattices employed in our analysis. In what
follows, we first briefly outline the basics of the method employed in 
Ref.~\cite{latticealphas}, then describe our implementation of this
approach, and finally present our results. We also discuss briefly
the differences (and complementarity) between our implementation
and that of the other recent similar, but not identical, 
reanalysis by HPQCD~\cite{hpqcd08}. An expanded discussion of the work
reported here may be found in Ref.~\cite{cssmalphas08}.

The authors of Ref.~\cite{latticealphas} extracted $\alpha_s(M_Z)$
by studying a large number of UV-sensitive
lattice observables, including the three, $\log(W_{11})$, 
$\log(W_{12})$ and $\log(W_{12}/u_0^6)$ on which we focus below.
The perturbative expansion for such an observable, $O_k$, is
written in the form
\begin{equation}
O_k=\sum_{N=1}d_{N}^{(k)}\hat{\alpha}(Q_k)^N
\equiv D_k\hat{\alpha}(Q_k)\sum_{M=0}c_M^{(k)}\hat{\alpha}(Q_k)^M
\label{3loopPT}\end{equation}
with $c_0^{(k)}\equiv 1$, $Q_k=d_k/a$ the Brodsky-Lepage-Mackenzie (BLM)
scale for the observable $O_k$, and $\hat{\alpha}$ any coupling
having the same expansion to $O(\alpha^3_s)$ (with $\alpha_s$ the
usual $\overline{MS}$ coupling) as the usual heavy quark potential 
coupling, denoted $\alpha_V^p$ below. The coefficients
$d_{1,2,3}^{(k)}$ (equivalently, $D_k, c_1^{(k)}$, and 
$c_2^{(k)}$) have been computed in 3-loop lattice perturbation 
theory~\cite{mason} for a number of such observables and, with 
the corresponding $d_k$, tabulated in Refs.~\cite{latticealphas,hpqcd08,mason}.
They are common for all such couplings $\hat{\alpha}$. The couplings 
$\hat{\alpha}$ also share common values for the first three $\beta$
function coefficients, $\hat{\beta}_0=9/4$, $\hat{\beta}_1=4$ and 
$\hat{\beta}_2=33.969$, where in our normalization
$\mu^2 \, d\hat{a}(\mu )/d\mu^2\ =\ -\sum_{n=0}\hat{\beta}_n \hat{a}^n(\mu )$
with $\hat{a}\equiv \hat{\alpha} /\pi$. When the expansion of $\hat{\alpha}$
is further specified to $O(\alpha_s^4)$,
$\hat{\beta}_3$ is also determined from the known values of the
4-loop $\overline{MS}$ $\beta$ function coefficients, 
$\beta_0,\, \cdots,\, \beta_3$~\cite{4loopbeta}.
The $\hat{\alpha}(Q_k)$ appearing in Eq.~(\ref{3loopPT}) are
determined by the value at a single reference scale, the
reference scale value serving as a fit parameter for the analysis.
It is, of course, important to remove any non-perturbative
contributions to the observable in question in order to make
use of Eq.~(\ref{3loopPT}). The reliability of the analysis
will be greatest when such non-perturbative subtractions are small.

Somewhat different choices for $\hat{\alpha}$ are made in the two
recent reanalyses. We denote our choice 
by $\alpha_T$, and that of Ref.~\cite{hpqcd08} by $\alpha_V$.
The relation between $\alpha_V^p$ and $\alpha_s$, to $O(\alpha_s^3)$, 
is of the form~\cite{schroder} 
\begin{equation}
\alpha_V^p(q^2)=\alpha_s(\mu^2)\left[1+\kappa_1(\mu^2/q^2)\alpha_s(\mu^2)
+\kappa_2(\mu^2/q^2)\alpha_s(\mu^2)\right]
\label{schrodereqn}\end{equation}
with the expressions for $\kappa_{1,2}(x)$ given in Ref.~\cite{schroder}.
The $n_f=3$ version of the RHS of Eq.~(\ref{schrodereqn}), with 
$\mu^2=q^2$, defines our ${\alpha}_{\, T}(q^2)$.
Numerically, 
\begin{equation}
\alpha_T(\mu^2 ) = \alpha_s(\mu^2 )\left[ 1+0.5570\alpha_s(\mu^2 )
+1.702\alpha^2_s(\mu^2 )\right]\ .
\label{alphavvsalphasnf3}\end{equation}
This exact (by definition) relation is used to run $\alpha_{\, T}$
between different scales using the intermediate $\alpha_s$ coupling, 
whose running can be reliably performed at 4-loops over the range of 
scales relevant to the observables considered. The HPQCD coupling, 
$\alpha_{\, V}$, is defined as follows. Beginning with Eq.~(\ref{schrodereqn}),
one takes the RHS, with $\mu^2=e^{-5/3}q^2$, to define an intermediate
coupling, $\alpha_{\, V}^\prime (q^2)$. This coupling has a $\beta$
function, $\beta^\prime$, with known values of 
$\beta^\prime_0,\cdots ,\beta^\prime_3$,
but also non-zero, but unknown, higher order coefficients,
$\beta^\prime_{4,5,\cdots}$, whose values depend on the
presently unknown $\beta_{4,5,\cdots}$.
The final HPQCD coupling, $\alpha_{\, V}$, is obtained from $\alpha_V^\prime$
by adding terms of $O(\alpha_s^5)$ and higher with coefficients chosen
in such a way as to make $\beta^V_4=\beta^V_5=\cdots =0$. Since
$\beta_{4,5,\cdots}$ are not known, the values of the coefficients
needed to implement these constraints are also not known.

Using the expansion parameter $\alpha_{\, T}$, 
no perturbative uncertainty is encountered in
converting the fitted reference scale $\alpha_{\, T}$
value to the equivalent reference scale $\overline{MS}$ result.
This is not true for the HPQCD parameter $\alpha_{\, V}$.
Higher order perturbative uncertainties, however, do remain in
our analysis. To see where these occur, 
and to understand the motivation for the alternate HPQCD
choice, let us define ${\alpha}_{\, 0}\equiv \hat{\alpha}(Q_0)$, 
where $Q_0$ is the maximum of the BLM scales
(corresponding to the finest lattice) for the observable in question.
We next expand the couplings at lower BLM scales (coarser
lattices) for the same observable, in the standard manner as a power
series in $\alpha_{\, 0}$,
\begin{equation}
\hat{\alpha}(Q_k)=\sum_{N=1}p_N(t_k)\alpha_0^N
\label{lowscaleexpansion}\end{equation}
where $t_k=\log\left( Q_k^2/Q_0^2\right)$, and the $p_N(t)$ are polynomials
in $t$ with coefficients determined by those of $\hat{\beta}$. Substituting
this representation into Eq.~(\ref{3loopPT}), one obtains the following
expression, where we replace any occurences of 
$\hat{\beta_0},\cdots ,\hat{\beta_2}$ 
with their known numerical values and display only those terms
involving one or more of $\hat{\beta_3}$ and
the unknown quantities $\hat{\beta_4},\hat{\beta_5},\cdots$,
$c_3^{(k)},c_4^{(k)},\cdots$:
\begin{eqnarray}
{\frac{O_k}{D_k}}\, &=&\, \cdots +
\alpha_0^4\left( c_3^{(k)}+ \cdots\right) + \alpha_0^5
\left( c_4^{(k)}-0.01027\hat{\beta}_3-2.865 c_3^{(k)} t_k+\cdots\right)
+ \alpha_0^6\left( c_5^{(k)} -0.00327 \hat{\beta}_4 t_k
\right. \nonumber\\
&&\left. \ \ \ \ \ 
-3.581 c_4^{(k)} t_k +[0.02573 t_k^2-0.02053c_1^{(k)}t_k]\hat{\beta}_3
[5.129 t_k^2 - 1.621 t_k] c_3^{(k)}+\cdots\right)
\nonumber\\
&&\ \ \  + \alpha_0^7 \left( c_6^{(k)}-0.001040 \hat{\beta}_5 t_k 
 +[0.009361 t_k^2-0.006536 c_1^{(k)} t_k] \hat{\beta}_4
\right.\nonumber\\
&&\left. \ \ \ \ \ 
+[-0.04213 t_k^3 +(0.01664+0.06617c_1^{(k)}) t_k^2-0.03080 c_2^{(k)} t_k] 
\hat{\beta}_3
-4.297c_5^{(k)}t_k \right.\nonumber\\
&&\left. \ \ \ \ \ +[7.694 t_k^2 - 2.026 t_k] c_4^{(k)}
+[- 7.347 t_k^3+ 6.386 t_k^2 -4.382 t_k] c_3^{(k)} +\cdots\right)+\cdots\ .
\label{cfitproblem}\end{eqnarray}
Running the $\overline{MS}$ coupling numerically using the
4-loop-truncated $\beta$ function is equivalent to keeping
terms involving $\beta_0,\cdots ,\beta_3$ to all orders,
and setting $\beta_4=\beta_5=\cdots =0$. The neglect of
$\beta_4,\beta_5\cdots$ also alters $\hat{\beta}_4,\hat{\beta}_5,\cdots$,
and hence produces a ``distortion'' of the true $t_k$-dependence, 
beginning at $O(\alpha_0^6)$. Since it is the scale-dependence of $O_k$ 
which allows one to fit the unknown coefficients $c_{3,4,\cdots}^{(k)}$, 
as well as $\alpha_0$, it follows that the 4-loop truncation 
forces compensating changes in at least the coefficients
$c^{(k)}_{4,5,\cdots}$. A shift in the values of $c_4^{(k)}$, however,
leads also to a shift in the $O(\alpha_0^5)$ coefficient which, in general,
will necessitate a compensating shift in $c_3^{(k)}$ as well. This in
turn will necessitate a shift in $\alpha_0$. Since the $\overline{MS}$
$\beta$ function is known only to 4-loop order, such 
truncated-running effects are unavoidable at some level. 
From Eq.~(\ref{cfitproblem}), however, it follows
that their size can be minimized by taking $Q_0$ as large as possible 
(achieved by working with the observable with the highest intrinsic BLM scale)
and keeping $t_k$ from becoming too large (achieved by restricting one's 
attention, if possible, to a subset of finer lattices). Note that,
by defining ${\alpha\,}_V$ in such a way that the 4-loop-truncated
$\beta$ function $\beta^V$ is exact, the HPQCD expansion parameter
choice, by definition, avoids these truncated-running problems. The price paid
is the unknown relation between $\alpha_{V}$ and $\alpha_s$
beyond $O(\alpha_s^4)$. The impact of this uncertainty
on $\alpha_s(M_Z)$ cannot be controlled, either through the choice of
observable or through the restriction to a subset of finer lattices.

From the discussion above we see that the two different coupling choices 
lead to complementary analyses. If the impact of the neglect
of higher order perturbative corrections in both cases is small,
the two approaches should give compatible results for analyses
based on the same observables, providing a form of mutual cross-check.
Good agreement is, indeed, found (see Ref.~\cite{cssmalphas08}
for details).

\section{Results}
We now turn to the results of our analysis. We employ data on our
observables from the MILC $a\sim 0.06$, $0.09$, $0.12$, $0.15$ and $0.18$ fm
ensembles~\cite{thanksto}. To minimize incompletely incorporated 
higher order perturbative contributions, we perform our main analysis 
using the three finest lattices, expanding to full 5-fold fits to test 
the stability of our solutions. The physical scales for the various
ensembles are determined using the measured values of $r_1/a$ and
the recent MILC assessment, $r_1=0.318(7)$~\cite{bernard07}.
The uncertainties on the extracted $\alpha_s$ values associated with
those on $r_1/a$ and $r_1$ are added linearly to arrive at
an ``overall scale uncertainty'' contribution to the total error. 

Quark-mass-dependent non-perturbative
contributions are found, using the data for different mass combinations
$am_\ell /am_s$, to be very linear in $2am_\ell +am_s$, allowing
these contributions to be fitted and removed with good reliability.
Because of details of the analysis not discussed here, the
uncertainty in this subtraction appears as part of ``the overall
scale uncertainty'' discussed above~\cite{cssmalphas08}.

Mass-independent non-perturbative contributions are assumed 
dominated by the $D=4$ gluon condensate contribution. The 
corresponding leading order contribution to the $m\times n$ Wilson 
loop $W_{mn}$, denoted by $\delta_gW_{mn}$, is known
from Ref.~\cite{wmnglue},
\begin{equation}
\delta_g W_{mn}\, =\, {\frac{-\pi^2}{36}}m^2n^2 a^4
\langle {\frac{\alpha_s}{\pi}}G^2\rangle\ .
\label{gluewmnterm}\end{equation}
As central input for the condensate we employ the result of the updated
charmonium sum rule analysis of Ref.~\cite{newgcond4},
$\langle {\frac{\alpha_s}{\pi}}G^2\rangle\, =\, (0.009\pm 0.007)\ {\rm GeV}^4$.
Since the error is already close to $100\%$,
we take the difference between results obtained
with and without the resulting subtraction as a measure of the
associated uncertainty. The neglect of mass-independent non-perturbative
contributions with $D>4$, for which no pre-existing constraints are available, 
should be safe so long as the estimated $D=4$ gluon condensate subtraction 
is small. 

The shifts associated with the central gluon condensate subtraction grow with 
increasing lattice spacing $a$ and, for the observables considered here, 
are (i) $-0.01\%$, $-0.02\%$ and $-0.08\%$ for $a\sim 0.06$ fm, 
(ii) $-0.1\%$, $-0.4\%$ and $-1.3\%$ for $a\sim 0.12$ fm, and 
(iii) $-0.5\%$, $-1.8\%$ and $-5.6\%$ for $a\sim 0.18$ fm, 
where in each case the values quoted correspond to the observables
$\log (W_{11})$, $\log (W_{12})$ and $\log (W_{12}/u_0^6)$, in that order.
The corrections, as claimed, are small, 
making the mass-independent non-perturbative subtraction safe for both
the central 3-fold and extended 5-fold fits~\cite{footnote}.
The subtractions are particularly small for the three finest lattices and 
for the plaquette observable, $\log (W_{11})$.

In line with the results of Ref.~\cite{latticealphas}, we find that,
even for the highest-scale observables and three finest lattices, 
the known terms in the perturbative expansion of the $O_k$ are insufficient 
to provide a description of the observed scale-dependence.
When $c_3^{(k)}$ is added to the fit, however, very good fit qualities
are found, with $\chi^2/dof<1$ (very significantly so
for the 3-fold fits). It follows that, with our current errors, 
it is not possible to sensibly fit additional coefficients
in the expansions of the $O_k$. This raises concerns about possible
associated truncation uncertainties. Since the relative weight of
higher order relative to lower order terms grows with decreasing
scale, the comparison of the results of the 3-fold and 5-fold fits
provides one handle on such a truncation uncertainty. If higher
order terms which have been neglected are in fact {\it not} negligible,
then the growth with decreasing scale of the resulting fractional error
should show up as an instability in the values of the parameters extracted
using the different fits. We see no signs for such an instability within the
errors of our fits, but nonetheless include a component equal
to the difference of central values obtained from the 3-fold and 5-fold
fits as part of our error estimate. This ``stability component'' is
added in quadrature with the overall scale uncertainty, the gluon
condensate subtraction uncertainty, and the small uncertainty 
associated with varying the $c_2^{(k)}$ (and, if relevant, $c_1^{(k)}$) within
the errors in their numerical evaluations to arrive at the total
error on our results.

After converting our result for the reference scale $n_f=3$ $\alpha_T$ 
coupling to the corresponding $n_f=3$ $\overline{MS}$ value, 
we run the result to $M_Z$ using the usual self-consistent combination
of 4-loop running and 3-loop matching~\cite{cks97}, taking
the flavor thresholds to lie at $rm_c(m_c)$ and $rm_b(m_b)$, with
$r$ varying between $1$ and $3$,
$m_c(m_c)=1.286\pm 0.013$ GeV and $m_b(m_b)=4.164\pm 0.025$ GeV~\cite{kss07}.
The evolution to $M_Z$ produces an additional $0.0003$ uncertainty
on $\alpha_s(M_Z)$~\cite{bck08}.

\begin{figure}
\unitlength1cm
\begin{minipage}[thb]{11.7cm}
\begin{picture}(11.6,13.1)
\epsfig{figure=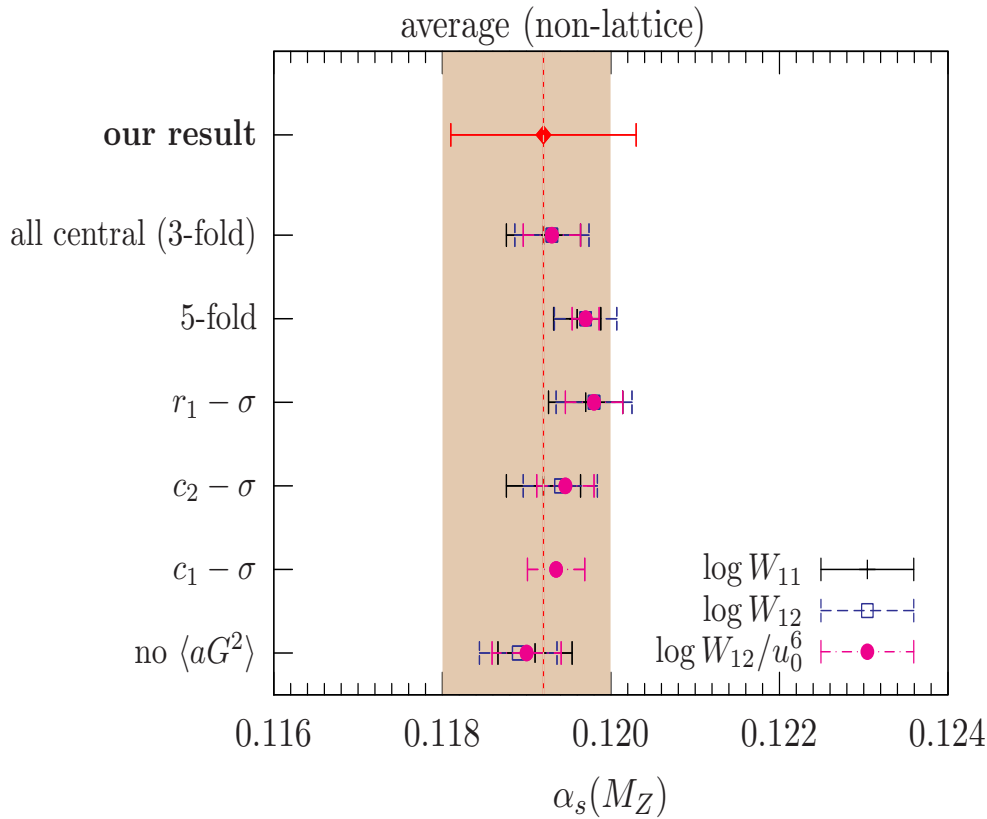,height=11.0cm,
width=13.0cm}
\end{picture}
\end{minipage}
\caption{Contributions to the errors on $\alpha_s(M_Z)$. Shown are
the results for $\alpha_s(M_Z)$ obtained using (i) the 3-fold fit
strategy, with central values for all input (the ``central'' case),
(ii) the alternate 5-fold fit strategy, still with central values
for all input, and (iii) the 3-fold fit strategy, with, one at a time,
each of the input quantities shifted from its central values by
$1\sigma$, retaining central values for the remaining inputs.
The error bars shown in each such case are those associated with the
uncertainties in $r_1/a$ for the various ensembles.}
\label{figure1}\end{figure}

Our results for $\alpha_s(M_Z)$, combining all errors as indicated above,
are $0.1192(11)$ from the fit based on $\log (W_{11})$, and $0.1193(11)$
from those based on $\log (W_{12})$ and $\log (W_{12}/u_0^6)$. The consistency
represents an improvement over that of the corresponding results quoted in
Ref.~\cite{latticealphas}. In line with the arguments above, we believe the
most reliable analysis to be that obtained using the three finest lattices,
and the observable, $\log (W_{11})$, having both the highest intrinsic scale 
and smallest gluon condensate subtractions. Our final determination,
based on this case, is thus
\begin{eqnarray}
\alpha_s(M_Z)=0.1192(11)\ .
\end{eqnarray}
The dominant contribution, $0.0009$, to the total error is that associated
with the overall scale uncertainty. A graphical depiction of the various
components of the error is given in Figure~\ref{figure1} where, for
clarity, only one-sided errors are shown. The results are in excellent
agreement with those of the independent determinations mentioned
above, whose average is shown in the figure by the shaded band.

\begin{thebibliography}{99}
\bibitem{footnote}The reader should bear in mind that the fractional $D=4$
gluon condensate corrections can be rather large for a number of the
Creutz ratios and tadpole-rescaled ratios considered in 
Ref.~\cite{hpqcd08} (see Ref.~\cite{cssmalphas08} for numerical details).
The HPQCD analysis in fact allows the gluon condensate
to be determined independently for each observable considered, using zero
central value and a width given by the conventional SVZ value
as priors in the full Bayesian fit. The variation in the values of
the gluon condensate across the different observables 
(as reported to us privately by the HPQCD collaboration) is 
likely to be due to the presence of neglected
mass-independent $D>4$ contributions. An update by the HPQCD collaboration
is in progress to incorporate such higher $D$ contributions into their 
fit form. The degree to which the output gluon condensate values corresponding
to different observables are in agreement will provide a useful test of the 
consistency of this approach.
\end{thebibliography}
\end{document}